# Estimation With Binned Data

Paul T. von Hippel
Igor Holas
Samuel V. Scarpino

*Abstract (147 words)*

Variables such as household income are sometimes *binned*—so that we only know how many households fall in each of several bins such as $0-10,000, $10,000-15,000, or $200,000+. We provide a SAS macro that estimates the mean and variance of binned data by fitting the extended generalized gamma (EGG) distribution, the power normal (PN) distribution, and a new distribution that we call the power logistic (PL). The macro also implements a "best-of-breed" estimator that chooses from among the EGG, PN, and PL estimates on the basis of likelihood and finite variance. We test the macro by estimating the mean family and household incomes of approximately 13,000 US school districts between 1970 and 2009. The estimates have negligible bias (0–2%) and a root mean squared error of just 3–6%. The estimates compare favorably with estimates obtained by fitting the Dagum, generalized beta (GB2), or logspline distributions.

Correspondence: Paul T. von Hippel is Assistant Professor, LBJ School of Public Affairs, University of Texas, 2315 Red River, Box *Y,* Austin, TX 78712, paulvonhippel.utaustin@gmail.com.

# 1   INTRODUCTION

Many social scientists take an interest in variables such as income, yet the distribution of such variables is often *interval-censored*. In interval censoring, rather than knowing the exact value of $X$—say household income—we instead know only that for household $i$, the value of $X_i$ falls within some *interval* $[m_i, M_i)$ where $m_i$ and $M_i$ are a minimum and maximum value. In principle each observation can have its own interval, but in practice the incomes of a large number of families or households are often summarized with a relatively small number of intervals or *bins*.

As an example of binned data, Table 1 summarizes census data on year 2000 household incomes within one of the richest and one of the poorest school districts in the United States: Rancho Santa Fe Elementary School District in California, and McNary Elementary School District in Arizona. For each district we know how many households $n_b$ have incomes in each of sixteen bins $[m_b, M_b)$, $b$=1,…,16. For example, in the poorest bin [$0,$10,000) there are 55 households in McNary and 45 households in Rancho Santa Fe, while in the richest bin [$200,000, +∞) there are 0 households in McNary and 910 households in Rancho Santa Fe.

←Table 1 near here→

It is clear that Rancho Santa Fe is much richer than McNary, but given only bins it is hard to be more specific. How many times larger is the average household income in Rancho Santa Fe than in McNary? How do the two districts compare with respect to the standard deviation of household income, or the coefficient of variation (the standard deviation divided by the mean)?

To answer these questions, we have to estimate the mean and variance from the bins. How can we do that? A simple approach is to assume that each family's income is at the *midpoint* of its bin. For example, we might assume that all households with incomes in the bin [$0,$10,000) have an income of exactly $5,000. This assumption is unrealistic, but it can be serviceable if the bins are narrow. If the bins are wide, then the midpoint approximation may be less accurate, since within some bins the distribution of households may be highly variable and may not be centered around the bin midpoint. The midpoint approximation also runs into practical difficulties if the data are "top-coded" so that the highest bin is unbounded or censored on the right—as in the Rancho Santa Fe school district, where nearly half the households are in the top bin [$200,000, +∞). Analysts commonly handle top-coding by assuming that the incomes within the top bin fit some distribution (e.g., Pareto). But such assumptions can be inaccurate and are hard to test (Hout 2004).

A more sophisticated approach is to fit a flexible distribution not just to the top bin, but to the entire distribution. The method of fitting should be appropriate for binned data, and the fitted distribution should be flexible enough to handle a variety of distributional shapes. In this paper, we estimate the mean and variance of income within each of approximately 13,000 school districts in various years from 1970 to 2009, using three distributions: the extended generalized gamma (EGG), the power normal (PN), and a new distribution that we call the power logistic (PL). We also propose a "best-of-breed" estimator that chooses from among the EGG, PN, and PL estimates the estimate that has the highest likelihood, provided that it also has a finite variance.

Our estimated means have a bias of 2% or less and a root mean squared error of 6% or less. The estimates compare quite favorably with alternative estimators obtained by fitting the Dagum, generalized beta (GB2), or logspline distributions.

We provide a SAS macro that implements our estimator. The use of the macro is described in the Appendix.

## 2   DATA

We use binned data on the incomes of households, families, and families with children within each of the approximately 13,000 school districts in the United States. The binned data are provided by the U.S. Bureau of the Census and the National Center for Education Statistics (National Center for Education Statistics 1970; Bureau of the Census 1983; National Center for Education Statistics 2012). The data are derived from the long-form Census until 2000 and the American Community Survey thereafter. We limit our evaluation to the years and income types where the Census Bureau provides both the district's "true" mean income and the distribution of incomes across bins: this means our analysis uses household incomes in 1980, 2000, and 2005-09, family income in 1970, 2000, and 2005-09, and the income of families with children in 2000. In fact, the "true" mean income is also an estimate, but the estimate is based on individual incomes that have not been binned (American Community Survey Office 2010), and so we regard these estimates as a reference which is as close to the truth as we are likely to get.

As shown in Table 1, the Census bin counts are rounded to the nearest 5, except for counts below 5 which are rounded to 4. This rounding can introduce substantial error in very small districts, so we eliminate from consideration any districts with fewer than 40 households, 40 families, or 40 families with children. We also eliminate any district where fewer than 4 bins have nonzero counts. With fewer than 4 nonzero bins, it would be difficult to fit a distribution with 3 or more parameters.

## 3   METHODS

Maximum likelihood offers a convenient way to estimate the distribution of binned data. Given $B$ bins and a cumulative distribution function $F(X|\boldsymbol{\theta})$ with a vector of parameters $\boldsymbol{\theta}$, the likelihood is

$$L(\boldsymbol{\theta}|data) = \prod_{b=1}^{B} P(m_b \leq X \leq M_b|\boldsymbol{\theta})^{n_b}$$

$$= \prod_{b=1}^{B} \left( F(M_b|\boldsymbol{\theta}) - F(m_b|\boldsymbol{\theta}) \right)^{n_b}$$

where $n_b$ is the number of households of families in the bin $[m_b, M_b]$, $b = 1, \ldots, B$. Then the maximum likelihood estimate $\widehat{\boldsymbol{\theta}}$ is the value of $\boldsymbol{\theta}$ that maximizes $L(\boldsymbol{\theta}|data)$. Summary statistics such as the mean and variance can then be estimated as functions of $\widehat{\boldsymbol{\theta}}$. It remains only to specify the cumulative distribution function $F(X|\boldsymbol{\theta})$ or equivalently the density $f(X|\boldsymbol{\theta})$. Given $f(X|\boldsymbol{\theta})$ we can derive formulas that relate $\boldsymbol{\theta}$ to the mean $E(X)$ and variance $V(X)$, or equivalently we can derive a formulas for the $k^{\text{th}}$ moment $E(X^k)$ since $V(X) = E(X^2) - \left( E(X) \right)^2$.

## 3.1   Our proposed method

A number of densities can be fit to binned data, including the one-parameter *exponential* density and two-parameter densities such as the *normal*, *logistic*, *lognormal*, *loglogistic*, *gamma*, and *Weibull*. But there is little point to fitting a one- or two-parameter density since they can usually be mimicked by a density with three parameters.

Our SAS macro *%fit_binned* fits three three-parameter distributions: the extended generalized gamma distribution (EGG), the power normal distribution (PN), and the power logistic distribution (PL). The macro can also choose the "best-of-breed" estimate on the basis of likelihood and finite variance. We discuss the computational details below.

### 3.1.1   Extended generalized gamma (EGG) distribution

The extended generalized gamma (EGG) distribution is a three-parameter distribution that includes as special cases the two-parameter normal, gamma, and Weibull distribution and the one-parameter exponential distribution. The density of the EGG distribution is

$$f(X) = \begin{cases} \dfrac{1}{\sigma X} \phi_{nor}(\omega) \text{ if } \lambda = 0 \\[2mm] \dfrac{|\lambda|}{\sigma X} \phi_{lg}(\lambda\omega + \ln(\lambda^{-2}), \lambda^{-2}) \text{ if } \lambda \neq 0 \end{cases} \tag{1}$$

where $X > 0$, $\sigma > 0$, $\omega = (\ln(X) - \mu)/\sigma$, $\phi_{nor}$ is the standard normal density, and $\phi_{lg}$ is the standard log-gamma density (Meeker and Escobar 1998).

The parameters $\mu, \sigma, \lambda$ of the EGG distribution can be estimated from binned data using the LIFEREG procedure in SAS. A limitation of the LIFEREG procedure is that it will not accept a bin with a lower bound of $m_b = 0$, and values close to $m_b = 0$ can cause the program to terminate due to numerical problems. Experimentation showed that these errors could be avoided by replacing $m_b = 0$ with $m_b = 1/2$ .

The $k^{\text{th}}$ moment of the EGG density can be expressed in terms of the parameters as

$$E(X^k) = \begin{cases} a. \ \exp\left(k\mu + \dfrac{1}{2}(k\sigma)^2\right) \text{ if } \lambda = 0 \\ b. \ +\infty \text{ if } k\lambda\sigma \leq -1 \\ c. \ \dfrac{\Gamma[\lambda^{-1}(k\sigma + \lambda^{-1})]}{\Gamma(\lambda^{-2})} \exp(k\mu) \ (\lambda^2)^{k\sigma/\lambda} \text{ if } \lambda \neq 0 \text{ and } k\lambda\sigma > -1 \end{cases} \qquad (2)$$

(Meeker and Escobar 1998). The possibility of infinite moments is a liability. Note that even if the mean (first moment) is finite, if the second moment is infinite the standard error of the first moment may be infinite as well. That is, the estimated mean may sometimes be extremely variable.

Formula (2)c can present numerical problems if $\lambda$ is close to zero, because then the $\Gamma$ functions can grow so large that they overflow a machine's floating-point representation. Such numerical problems can usually be avoided by expressing $E(X^k)$ in terms of $ln\Gamma$ instead of $\Gamma$. (SAS, like some other languages, implements $ln\Gamma$ as a separate function that does not require prior calculation of $\Gamma$.) So if $\lambda \neq 0$ and $k\lambda\sigma \leq -1$, then

$$E(X^k) = \exp\left(k\mu + \frac{k\sigma}{\lambda}ln(\lambda^2) + ln\Gamma[\lambda^{-1}(k\sigma + \lambda^{-1})] - ln\Gamma(\lambda^{-2})\right) \qquad \begin{array}{c}(2)c, \\ \text{revised}\end{array}$$

If $\lambda$ is very close to zero it is possible for even the $ln\Gamma$ functions to overflow the computer's floating-point representation. We did not encounter this situation in practice, but it seems wise to plan for it. If $ln\Gamma$ overflows, then we can use Formula (2)a, which is exact for $\lambda = 0$ and approximate for $\lambda$ near zero. We can also improve on Formula (2)a by noting that near $\lambda = 0$, $E(X^k)$ is approximately linear in $\lambda$. Empirically, near $\lambda = 0$ the following approximations work well for the first two moments of the school-district data:

$$\begin{aligned} E(X) &= E(X|\lambda = 0) + \lambda/2 \\ E(X^2) &= E(X^2|\lambda = 0) + 3\lambda/2 \end{aligned} \qquad \begin{array}{c}(2)a, \\ \text{revised}\end{array}$$

### 3.1.2   Power-normal (PN) distribution

The power-normal (PN) distribution assumes that a nonnegative variable $X$ can be transformed to approximate normality by a power transformation with power $\lambda$

$$t(X, \lambda) = \begin{cases} X^\lambda \text{ if } \lambda \neq 0 \\ \ln(X) \text{ if } \lambda = 0 \end{cases} \qquad (3)$$

so that the transformed variable $t(X, \lambda) \sim N(\mu, \sigma)$ follows a normal distribution (Box and Cox 1964; Freeman and Modarres 2006).[1] This is somewhat unrealistic since if $\lambda \neq 0$, $t(X, \lambda)$ is nonnegative and cannot be exactly normal. Nevertheless, $t(X, \lambda)$ can be quite close to normal even if $X$ has substantial density near zero (Hawkins and Wixley 1986). In fact, with $\lambda = 1/4$ the PN distribution can mimic very accurately the two-parameter gamma distribution (Hawkins and Wixley 1986). The PN distribution can also mimic the Weibull distribution, since a power transformation can convert the Weibull into an exponential distribution (Keats, Nahar, and Korbel 2000), and the exponential distribution is a special case of the gamma distribution.

---

[1] There are versions of the power-normal distribution where $t(X,\lambda)$ is modified to ensure continuity at $\lambda=0$ (e.g., Box and Cox 1964). For our purposes this is unnecessary and would only complicate the calculations.

If we take the approximation literally, then we can invert the power transformation to express $X$ as a transformation of a truly normal variable $Z \sim N(\mu, \sigma)$.

$$X = t^{-1}(Z, \lambda) = \begin{cases} Z^{\lambda^{-1}} \text{ if } \lambda \neq 0 \\ exp(Z) \text{ if } \lambda = 0 \end{cases} \tag{4}$$

Note that if $\lambda \neq 0$ we must assume that $\lambda^{-1}$ is a positive integer; otherwise $t^{-1}(Z, \lambda)$ is not a real number when $Z < 0$. So $\lambda$ can be $\frac{1}{2}, \frac{1}{3}, \frac{1}{4}$, etc.—i.e., $Z$ can be the square, cube, or fourth root of $X$—but $\lambda$ cannot be .26, for example, since $.26^{-1}$ is not an integer.

The moments of $X$ can be expressed in terms of the parameters $\mu, \sigma, \lambda$. If $\lambda = 0$ then $X$ is lognormal with moments

$$E(X^k) = \exp\left(k\mu + \frac{k^2\sigma^2}{2}\right) \tag{5}$$

If $\lambda^{-1}$ is a positive integer, the $k^{\text{th}}$ moment of $X$ is simply the $k/\lambda^{\text{th}}$ moment of the normal variable $Z$, which is readily calculated using Mathematica software, version 8:

$$E(X^k) = E(Z^{k/\lambda}) = \begin{cases} \sigma^{k/\lambda}\left(\frac{k}{\lambda} - 1\right)!! \, _1F_1\left(-\frac{k}{2\lambda}; \frac{1}{2}; -\frac{\mu^2}{2\sigma^2}\right) \text{ if } k/\lambda \text{ is even} \\ \mu\sigma^{k/\lambda-1}\frac{k}{\lambda}!! \, _1F_1\left(\frac{1}{2}\left(1 - \frac{k}{\lambda}\right); \frac{3}{2}; -\frac{\mu^2}{2\sigma^2}\right) \text{ if } k/\lambda \text{ is odd} \end{cases} \tag{6}$$

Here !! is the double-factorial function and $_1F_1$ is the confluent hypergeometric function.

Numerical implementation of formula (6) presents some challenges. For one thing, there is no single numerical recipe that provides the correct value of $_1F_1$ for all values of $k, \mu, \sigma, \lambda$ (Muller 2001). Fortunately, for practical purposes we do not need $E(X^k)$ for all values of $k$ and $\lambda$; it suffices to have the first two moments $E(X^k)$, $k = 1, 2$, for a reasonable variety of positive integers $\lambda^{-1}$. Even more fortunately, for $k = 1, 2$, and a particular positive integer $\lambda^{-1}$, formula (6) reduces to a more easily evaluated polynomial with degree $k\lambda^{-1}$. For example, with $\lambda^{-1} = 1$, $X$ is normal and the first two moments reduce to the polynomials $E(X) = \mu$ and $E(X^2) = \mu^2 + \sigma^2$. Likewise with $\lambda^{-1} = 2$, $X$ is noncentral chi-square and the first two moments reduce to $E(X) = \mu^2 + \sigma^2$ and $E(X^2) = \mu^4 + 6\mu^2\sigma^2 + 3\sigma^4$. Polynomials for the mean and variance with $\lambda^{-1} = 1, 2, ..., 20, 25, 33, 50$ are incorporated into our SAS macros.[2]

---

[2] Initial implementations included $\lambda^{-1} = 100$ as well, but the resulting $100^{\text{th}}$- and $200^{\text{th}}$-degree polynomials sometimes presented numerical problems.

For a given value of $\lambda$, fitting the PN distribution is a matter of transforming the bin endpoints[3] $[m_b, M_b]$ to $[t(m_b, \lambda), t(M_b, \lambda)]$ then fitting a normal model $X \sim N(\mu, \sigma)$ to the transformed bins using software that can fit normal interval-censored data, such as the LIFEREG procedure in SAS. We try different values of $\lambda$ and choose the one with the greatest model likelihood. Finally, we plug the maximum likelihood estimates[4] $\hat{\lambda}, \hat{\mu}, \hat{\sigma}$ into polynomials derived from equation (6) to obtain maximum likelihood estimates of the mean and variance.

### 3.1.3    Power-logistic (PL) distribution

The power-logistic (PL) distribution is a new distribution, as far as we know. It is just like the power-normal distribution except that $Z = t(X, \lambda) \sim Logistic(\mu, \sigma)$ has a logistic distribution rather than a normal distribution. Since the logistic distribution has heavier tails than the normal distribution, it follows that the PL distribution $X = t^{-1}(Z, \lambda)$ will have a somewhat heavier right tail than the PN distribution. This may be useful for heavy-tailed variables such as income.

As with the PN distribution, with the PL distribution we require that either $\lambda = 0$ or $\lambda^{-1}$ is a positive integer. The PL distribution is fitted like the PN distribution. We transform the bin endpoints $[m_b, M_b]$ to $[t(m_b, \lambda), t(M_b, \lambda))$ then fit a logistic model $X \sim Logistic(\mu, \sigma)$ to the transformed bins using the LIFEREG procedure in SAS.

As with the PN distribution, with the PL distribution the moments of $X$ can be expressed in terms of the parameters $\mu, \sigma, \lambda$. If $\lambda = 0$ then $X$ is log-logistic with moments

$$E(X^k) = \frac{\sigma^k}{\text{sinc}\left(\frac{k\pi}{\mu}\right)} \tag{7}$$

if $k < \mu$; otherwise $E(X^k)$ is indeterminate. The possibility of indeterminate higher moments is a liability of the log-logistic distribution, but a natural consequence of its heavy tail. We note that if the variance is indeterminate, the mean may be highly variable since the central limit theorem does not guarantee a finite standard error. In our SAS implementation, we exclude estimates that result in an indeterminate variance. This means that $\lambda = 0$ (the log-logistic distribution) will not be used if $\mu < 2$.

If $\lambda^{-1}$ is a positive integer, the $k^{\text{th}}$ moment of $X$ is simply the $k/\lambda^{\text{th}}$ moment of the logistic variable $Z$, which again is readily calculated using Mathematica software:

---

[3] If $m_b = 0$ and a log transformation is used, then $\ln(m_b) = -\infty$, which is coded as a missing value in SAS. The LIFEREG procedure can also fit the log-normal distribution directly, but SAS's implementation will ignore any bin with $m_b = 0$.

[4] In our implementation, the precision of the maximum likelihood estimate $\hat{\lambda}$ is limited because there are some values of $\lambda$ whose likelihood is not estimated. The values considered are limited to $\lambda^{-1} = 1,2,\ldots,20,25,33,50$ but it is possible that the exact maximum likelihood estimate could be an in-between value such as $\lambda^{-1} = 30$. This degree of imprecision is unimportant because for purposes of estimating the mean and variance there are only trivial differences between an estimate of, say, $\lambda^{-1} = 30$ and $\lambda^{-1} = 33$. The estimates $\hat{\mu}, \hat{\sigma}$ are maximum likelihood estimates when conditioned on $\hat{\lambda}$.

$$E(X^k) = E(Z^{k/\lambda}) = (2\pi)^{k/\lambda}(-i\sigma)^{k/\lambda}B_{k/\lambda}\left(\frac{i\mu}{2\pi\sigma} + \frac{1}{2}\right) \tag{8}$$

where $B_{k/\lambda}$ is the $k/\lambda^{\text{th}}$ Bernoulli polynomial. The presence of Bernoulli polynomials and imaginary numbers may be inconvenient for purposes of implementation, but again we can avoid the inconvenience by noting that we only need to calculate $E(X^k)$ for $k = 1,2$ and selected positive integers $\lambda^{-1} \leq 50$. For each particular value of $k$ and $\lambda^{-1}$, $E(X^k)$ simplifies to a polynomial of degree $k/\lambda$. For example, with $\lambda^{-1} = 1$, $X$ is logistic and the first two moments are given by the polynomials $E(X) = \mu$ and $E(X^2) = \mu^2 + \pi^2\sigma^2/3$. Polynomials for $k = 1,2$ and $\lambda^{-1} = 1,2,..,20,25,50$ are incorporated into our SAS macros.

### 3.1.4  Best-of-breed estimator

Once EGG, PN, and PLL estimates have been obtained, we can eliminate any estimates that return an indefinite or infinite variance and then choose among the remaining estimates the one that has the highest (log) likelihood. The estimate chosen in this way is called the "best of breed."

We require the best-of-breed estimate to have finite variance because an indefinite variance estimate is not useful, and because an indefinite variance estimate is often a sign that the mean estimate, if it is finite, is highly variable. What the finite-variance requirement means in practice is that we occasionally eliminate the EGG estimate from contention; the PN estimate is always in contention because it always has a finite variance; and the PL estimate is always in contention because, although the PL distribution can have an infinite variance if $\hat{\lambda} = 0$, we eliminated such infinite-variance estimates when we defined the PL estimator in Section 3.1.3.

## 3.2  Other methods

In principle, our best-of-breed approach could be extended to a larger breed of distributions, beyond the EGG, PN, and PL. We could choose among additional three-parameter distributions using the likelihood criterion, and we could choose among distribution with more parameters by using the Akaike Information Criterion or the Bayes Information Criterion, which penalizes the likelihood for extra parameters (Akaike 1974; Schwarz 1978).

We limited the *%fit_binned* macro to the EGG, PN, and PL distributions because they are the only three distributions that we could fit conveniently using the LIFEREG procedure in SAS. The LIFEREG procedure does fit other distributions, including the Weibull, exponential, and two-parameter gamma, but these are all special cases of the EGG distribution.

Below we discuss alternative distributions that can be fit using different software.

### 3.2.1 Dagum distribution

The Dagum distribution is a heavy-tailed distribution that was developed explicitly to model income (Kleiber 2008). Its density has three positive parameters $a, b, p$ and is defined for positive $X$:

$$f(X) = \frac{X^{ap-1} ap \left( \left( \frac{X}{b} \right)^a + 1 \right)^{-p-1}}{b^{ap}} \tag{9}$$

The moments of the Dagum density can be expressed in terms of the parameters as

$$E(X^k) = b^k \frac{\Gamma\left(1 - \frac{k}{a}\right) \Gamma\left(\frac{k}{a} + p\right)}{\Gamma(p)} \tag{10}$$

if $k < a$; otherwise $E(X^k)$ is indeterminate. The possibility of indeterminate higher moments is a liability of the Dagum distribution, but a natural consequence of its heavy tail.

The Dagum distribution can be fit to binned or interval-censored data using the *dagfit* module for Stata (Nichols 2010a). We are not aware of a way to fit the same model in SAS.

### 3.2.2 Generalized beta (GB2) distribution

The generalized beta distribution of the second kind (GB2) is a flexible four-parameter distribution that includes as special cases the three-parameter Dagum distribution and a three-parameter gamma distribution that is slightly different from the EGG distribution (Bandourian, McDonald, and Turley 2002). The GB2 density has three positive parameters $b, p, q$ and one additional parameter $a$ that can be positive or negative. It is defined for positive $X$:

$$f(X) = X^{ap-1} |a| \frac{\left( 1 - \left( \frac{X}{b} \right)^a \right)^{q-1}}{b^{ap} \mathrm{B}(p, q)} \tag{11}$$

where B is the beta function. The moments of the GB2 density can be expressed in terms of the parameters as

$$E(X^k) = b^k \frac{\mathrm{B}\left(p + \frac{k}{a}, q - \frac{k}{a}\right)}{\mathrm{B}(p, q)} \tag{12}$$

if $k < a$; otherwise $E(X^k)$ is indeterminate. The possibility of indeterminate moments is a liability of the GB2 distribution, but a natural consequence of its potential for heavy tails.

The GB2 distribution can be fit to binned or interval-censored data using the *gbgfit* module for Stata (Nichols 2010b). We are not aware of a way to fit the same model in SAS.

### 3.2.3   Logspline density estimation

A very general approach to fitting binned data is *logspline* density estimation, which merely assumes that the log of the density, i.e., $ln(f(X|\boldsymbol{\theta}))$, is a smooth cubic spline (Kooperberg and Stone 1991). The logspline model can fit distributions that are unimodal or multimodal and positively or negatively skewed, which makes it more flexible than the other distributions in this paper, all of which assume that the distribution is unimodal and the skew is nonnegative.

Logspline estimation may be viewed as a nonparametric technique in the sense that it makes few assumptions about the distribution, but the density fit by the logspline method does have parameters $\boldsymbol{\theta}$, namely the coefficients of the cubic polynomial and the positions of the knots. Once the number of knots has been decided, $\boldsymbol{\theta}$ can be estimated by maximum likelihood.

Some parametric densities may be viewed as special cases of the logspline density. For example, if $f(X|\boldsymbol{\theta})$ is normal then $ln(f(X|\boldsymbol{\theta}))$ is a quadratic with no knots. Simulations show that a logspline density can also mimic a gamma distribution or a mixture of a point mass with a smooth density (Kooperberg and Stone 1992). The flexibility of the logspline method gives it potential for overfitting, but this potential has been reduced by penalizing the method for extra parameters (Kooperberg and Stone 1992).

Logspline density estimation is implemented as part of the *polspline* package in R version 2.14 (Kooperberg 2010). Within that package, the *logspline* command fits uncensored data and the *oldlogspline* command fits interval-censored data. The *oldlogspline* command relies on legacy C code that has not been updated for 20 years. As might be expected given its age, the *oldlogspline* function has some technical issues. Most seriously, when the model fails to converge, the command returns a fatal exception error, which results in the loss of all previously obtained estimates. If we were fitting a single school district, we could work around a fatal exception by re-running the command with different initial parameter settings, but such an approach is not viable when we are looping the command over approximately 13,000 school districts. We found that we could avoid fatal exceptions by fixing the number of knots at 3. Note that only one of the three knot positions is actually a freely estimated parameter; the others are exterior knots fixed at the minimum value of the largest and smallest bin (i.e., $m_1, m_B$). For example, for household income in 2000 the exterior knots are at $0 and $200,000. In addition to its one interior knot, the smooth cubic spline model has 5 polynomial coefficients[5], so the logspline model has 6 free parameters. This is a lot of parameters for data with 16 bins, especially when some of the bins are empty or have very low counts (Table 1). We might therefore expect some difficulties with non-unique solutions or overfitting.

Past research has sometimes logged income before fitting a logspline density, though a logspline density can also be fit without first logging income (Kooperberg and Stone 1992). In our evaluation, we try both possibilities.

We do not know a general method for estimating the moments of a logspline density from its parameters $\boldsymbol{\theta}$, so instead we averaged 100 evenly spaced quantiles—namely quantiles .005, .015, .025,…,.995. These quantiles are calculated by the *qoldlogspline function in* the *polspline* package. For a heavy tailed distribution, the package sometimes returns an infinite or undefined value (Inf or NA) for the highest quantiles; under those circumstances, we take the mean and variance to be undefined as well.

---

[5] There would be more coefficients except that the spline is required to be continuously differentiable at the knot.

Our use of the logspline method is somewhat novel. Past applications have typically treated the method in an exploratory fashion, trying various initial parameter settings and choosing among the results after visual inspection. Ours is the first attempt to use the method in a wholly automatic fashion, running it over all U.S. school districts without manual intervention.

### 3.2.4 Histospline method

In addition to the logspline method, we also considered the similar *histospline* method (Wahba 1976), which fits a spline to a histogram. The histospline method is implemented in the *bda* package for R (Wang 2012), but it is not appropriate for our data because it assumes all bins are equal in width and bounded on both sides.

# 4 RESULTS

We evaluate the different methods with respect to their accuracy in estimating the mean income $E_d = E(X_d)$ of each school district $d$.[6] We concentrate on the mean rather than the variance because the Census reports the true mean but does not report the true variance for each school district. We begin with detailed results for household incomes in 2000, then broaden the evaluation to cover all years and all income types.

## 4.1 Household incomes in 2000

Figure 1 displays logarithmic scatterplots in which the estimate of mean family income $\hat{E}_d$ is plotted against the true means $E_d$. If a model estimated the mean with no error, all the points would fall on the diagonal line $\hat{E}_d = E_d$, which we show for reference. Where estimates tend to fall below the line, the estimator is negatively biased; where estimates tend to fall above the line, the estimator is positively biased. Figure 2 shows the biases explicitly by plotting the *relative error* $(\hat{E}_d - E_d)/E_d$—that is, the error as a percentage of the true mean—with a local regression curve showing whether the expected error is positive or negative in rich or poor districts.

It is evident from the scatterplots that the EGG and PN estimates produce very similar estimates, and share a slight negative bias in the richest districts. The PL estimates, by contrast, have a positive bias in the richest 1% of districts. The best-of-breed estimates are nearly unbiased because they mix the positively biased PL estimates with the negatively biased PN and EGG estimates.

---

[6] We use the symbol $E$ instead of $\mu$, because the Methods section used $\mu$ for other purposes.

Different distributions have advantages in different years and for different types of income. For example, for household income, the PL distribution is best of breed for most districts in 2000, but the EGG distribution is best of breed for a plurality of districts in 1980 and 2005-09.

For the fitted PN and PL distributions, the most common exponents are $\hat{\lambda} = 1/3$ and $\hat{\lambda} = 1/4$. Since a PN distribution with these exponents is very similar to a gamma distribution (Hawkins and Wixley 1986), it is perhaps unsurprising that the PN and EGG models yield such similar estimates. The disadvantage of the EGG distribution is that it can yield an undefined variance estimate, whereas the variance estimated under the PN distribution is always finite. In addition, in 2–5% of districts the EGG distribution either fails to converge or issues a warning on the way to convergence. In the event of a convergence failure or warning, however, the EGG distribution still provides estimates, and the estimates appear to be as accurate as they are under other circumstances. In fact, the likelihood for an EGG distribution that failed to converge is sometimes higher than the competing likelihoods of the PN and PL distributions.

Figure 3 shows parallel results for the Dagum and GB2 distribution. The figure shows that the Dagum estimates are nearly unbiased except for a slight positive bias in the richest 1% of districts. What is not as clear from the figure is that the Dagum estimates are a bit more variable than the EGG, PN, PL, or best-of-breed estimates. The Dagum estimates are therefore less accurate despite being nearly unbiased.

The GB2 estimates are inferior to the Dagum estimates. The GB2 estimates have a strong negative bias in the richest districts; in fact, in 18% of districts, the GB2 estimate of the mean is undefined and does not appear in the plot. In addition, in 37% of districts, the GB2 estimate of the variance is undefined. Many of the rich districts with a negatively biased mean also have an undefined variance.

It at first seems surprising that the GB2 estimates are worse than the Dagum estimates, because the three-parameter Dagum distribution is just a special case of the four-parameter GB2 distribution. Evidently the extra parameter in the GB2 distribution is not necessarily an asset. One possible explanation for this is that the fourth parameter opens up parts of the parameter space where the mean or variance is undefined. Another possible explanation is that four parameters may be too many for a distribution that is specified by only 16 bins.

Figure 4 summarizes the estimates that result from fitting a logspline density to household income or to the log of household income. As under the GB2 estimates, under the logspline estimates we find that extra flexibility does not bring better estimates. The logspline estimates display a substantial negative bias in the richest districts, and if income is logged the logspline estimates are negatively biased in the poor and middle-class districts as well. In addition, if income is logged, 63% of districts have an undefined mean or variance; naturally the undefined means do not appear in the plot. Undefined moments are much rarer if income is not logged.

## 4.2 All years, all income types

Table 2 summarizes the results for all the estimators, every income type, and every year. Estimators are compared with respect to several criteria: the relative bias, which is the mean of the relative errors; the root mean square of the relative errors (RMSRE), and the percentage of districts for which the estimator returns an undefined mean or variance.

For every year and income type, we find that the best estimates come from the EGG, PN, PL, and best-of-breed models. Those estimators have absolute biases of 2% or less and RMSREs of 6% or less, and they always yield a definite mean and variance, except for the EGG model which produces an undefined mean and variance in less than 1% of all districts. The best-of-breed estimates are not always better than the estimates obtained from the EGG, PN, or PL models individually. In fact, all the differences among the estimators are fairly small with respect to overall RMSRE or relative bias—although, as Figure 1 and Figure 2 show, the EGG, PN, and PL estimators, despite their general similarity, can still have noteworthy differences and biases in a few districts at the high or low end of the income distribution.

Among the other estimators, the best is the Dagum model, which has negligible bias but greater RMSRE than the EGG, PN, PL, and best-of-breed estimators. After the Dagum estimates, the next best estimates, in order, come from the logspline model fitted to income, the GB2 model, and the logspline model fitted to the log of income. For the GB2 model and the logspline model of logged income, a substantial liability is that the mean and variance are frequently undefined.

For some years and income types, the estimated bias and RMSRE of selected models can be extremely large, exceeding 100,000%. This occurs when the logspline model is fit to the log family income and when the Dagum model is fit to household income in 2005-09. The reason is that the models produce extremely large estimates for one or more districts; this occurs when the model parameters get close to a region where the mean would be infinite or undefined.

It is worth discussing past research that has reported better luck fitting some of these distributions to income data. Bandourian et al. (2002) fit the Dagum and GB2 distributions to binned household income data within 23 developed countries, without any examples of an undefined mean or variance. Kooperberg and Stone (1992) fit a logspline density to the log of British household incomes; the fitted density looks quite plausible, although it is not clear how the data were censored or whether the mean and variance of the fitted density were defined.

There are important differences between the circumstances of previous studies and our own. We fit models to approximately 13,000 different school districts, and in doing so we inevitably encountered some challenging circumstances that one would not encounter in fitting models to the income distribution of a single country, or even 23 countries. An important challenge occurs when the bulk of the income distribution is concentrated in a small number of bins; Table 1 gives two examples, one where half the incomes are in the lowest bin, and one where nearly half the incomes are in the highest bin. These circumstances did not arise in Bandourian et al.'s (2002) study, where each country's incomes were grouped into 20 "equal-probability bins," each containing 5% of households. It seems reasonable to suppose that a distribution with 20 equal-probability bins would be easier to fit, and would accommodate a model with more parameters, than a distribution where just a handful of bins contain most of the information that is relevant to estimation. A final difference is that Bandourian et al. (2002) fit the GB2 and Dagum their own Matlab implementation; they did not use the Stata commands that are tested in our paper.

Finally, we should acknowledge that our study is limited to comparing models' accuracy in estimating the district mean. The mean is a limited criterion, and it may be that certain models do a less-than-optimal job of estimating the mean and yet do a superior job of estimating the shape of the distribution or estimating shape-related quantities such as the percentiles or the mode. On the other hand, if a model does a good job of fitting a distribution's shape, we would expect it to provide estimates of the mean that, if not optimal, are at least fairly good. On this basis, it seems more plausible that good estimates of the percentiles or mode could be obtained using a model like the Dagum distribution of family income in 1970, whose mean estimates are approximately unbiased and only a little more variable than estimates obtained from the EGG, PN, or PL models. By contrast, it seems unlikely that the logspline method, which provides seriously biased mean estimates, would provide good estimates of shape.

# 5 CONCLUSION

Our results highlight the virtues of simplicity. We obtained inaccurate estimates using the flexible six-parameter logspline model or a four-parameter GB2 model, while we obtained relatively accurate estimates using simpler three-parameter models such as the EGG, PN, and PL. Among the three-parameter models, the Dagum model produced the least accurate estimates.

We have implemented the EGG, PN, and PL models in a SAS macro called *%fit_binned*, which can also choose among those models to provide a best-of-breed estimates. The use of the *%fit_binned* macro is described in the Appendix.

# TABLES

Table 1. Distribution of 2000 household income in two U.S. school districts.

| Income bins | | Number of households | |
|---|---|---|---|
| Min | Max | McNary | Rancho Santa Fe |
| $0 | $10,000 | 55 | 45 |
| $10,000 | $15,000 | 15 | 40 |
| $15,000 | $20,000 | 10 | 50 |
| $20,000 | $25,000 | 0 | 25 |
| $25,000 | $30,000 | 10 | 25 |
| $30,000 | $35,000 | 4 | 55 |
| $35,000 | $40,000 | 4 | 20 |
| $40,000 | $45,000 | 0 | 30 |
| $45,000 | $50,000 | 4 | 20 |
| $50,000 | $60,000 | 4 | 55 |
| $60,000 | $75,000 | 0 | 85 |
| $75,000 | $100,000 | 4 | 135 |
| $100,000 | $125,000 | 0 | 175 |
| $125,000 | $150,000 | 0 | 100 |
| $150,000 | $200,000 | 0 | 155 |
| $200,000 | $\infty$ | 0 | 910 |
| Total | | 110 | 1,925 |

Table 2. Accuracy of competing estimators for mean income

| Income | Year | N | | EGG | PN | PL | Best of breed | Dagum | GB2 | Logspline Income | Logspline Logged income |
|---|---|---|---|---|---|---|---|---|---|---|---|
| Families with children | 2000 | 13,681 | RMSRE | 6% | 5% | 6% | 6% | 9% | 11% | 12% | 21% |
| | | | Relative bias | -1% | -1% | 1% | 0% | 0% | -5% | -6% | -18% |
| | | | Undefined means | 0.4% | 0% | 0% | 0% | 0.1% | 17% | 0.6% | 51% |
| | | | Undefined variance | 0.4% | 0% | 0% | 0% | 0.5% | 52% | 0.6% | 51% |
| Families | 2000 | 14,024 | RMSRE | 5% | 5% | 5% | 5% | 8% | 10% | 12% | $4\times10^{100}$% |
| | | | Relative bias | -2% | -2% | 0% | -1% | 0% | -6% | -8% | $5\times10^{98}$% |
| | | | Undefined means | 0.2% | 0% | 0% | 0% | 0.0% | 16% | 0.3% | 49% |
| | | | Undefined variance | 0.2% | 0% | 0% | 0% | 0.3% | 42% | 0.3% | 49% |
| | 2005-09 | 13,592 | RMSRE | 5% | 5% | 5% | 5% | 11% | 15% | 14% | $7\times10^{98}$% |
| | | | Relative bias | -2% | -2% | 0% | -2% | 0% | -9% | -10% | $8\times10^{96}$% |
| | | | Undefined means | 0.6% | 0% | 0% | 0% | 0.3% | 43% | 1.6% | 55% |
| | | | Undefined variance | 0.6% | 0% | 0% | 0% | 1.0% | 67% | 1.6% | 55% |
| | 1970 | 12,361 | RMSRE | 3% | 3% | 3% | 3% | 4% | 5% | 6% | $5\times10^{96}$% |
| | | | Relative bias | 0% | 0% | 1% | 1% | 0% | -2% | -4% | $4\times10^{94}$% |
| | | | Undefined means | 0.02% | 0% | 0% | 0% | 0.01% | 3% | 0.0% | 7% |
| | | | Undefined variance | 0.02% | 0% | 0% | 0% | 0.07% | 64% | 0.0% | 7% |
| Households | 2000 | 14,120 | RMSRE | 5% | 5% | 5% | 5% | 7% | 10% | 11% | 26% |
| | | | Relative bias | -2% | -2% | 0% | -1% | 0% | -6% | -8% | -22% |
| | | | Undefined means | 0.2% | 0% | 0% | 0% | 0.0% | 18% | 0.04% | 63% |
| | | | Undefined variance | 0.2% | 0% | 0% | 0% | 0.4% | 37% | 0.04% | 63% |
| | 2005-09 | 14,120 | RMSRE | 5% | 4% | 5% | 5% | 487,599% | 14% | 14% | 31% |
| | | | Relative bias | -2% | -2% | 1% | -1% | 4,175% | -9% | -10% | -28% |
| | | | Undefined means | 0.3% | 0% | 0% | 0% | 0.1% | 50% | 0.3% | 83% |
| | | | Undefined variance | 0.3% | 0% | 0% | 0% | 0.9% | 71% | 0.3% | 83% |
| | 1980 | 13,668 | RMSRE | 3% | 3% | 5% | 3% | 17% | 7% | 7% | 21% |
| | | | Relative bias | 0% | 0% | 2% | 0% | 1% | -4% | -5% | -19% |
| | | | Undefined means | 0.1% | 0% | 0% | 0% | 0.1% | 21% | 0.2% | 63% |
| | | | Undefined variance | 0.1% | 0% | 0% | 0% | 0.3% | 49% | 0.2% | 63% |

# FIGURES

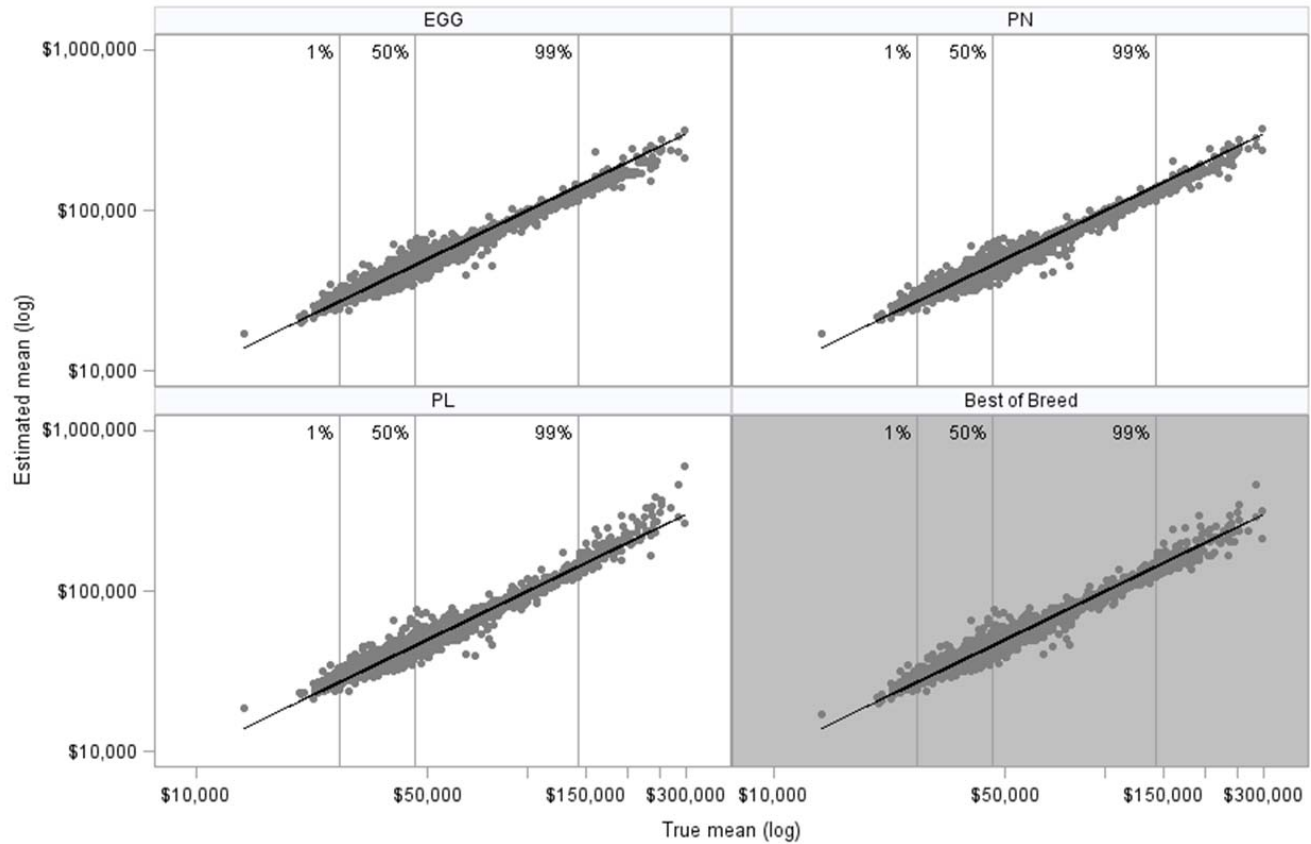

Figure 1. Mean household income for U.S. school districts in 2000. True means vs. means estimated under the EGG, PN, PL, and best-of-breed models. Perfectly accurate estimates would follow the diagonal line. The vertical lines are the 1[st], 50[th], and 99[th] percentiles.

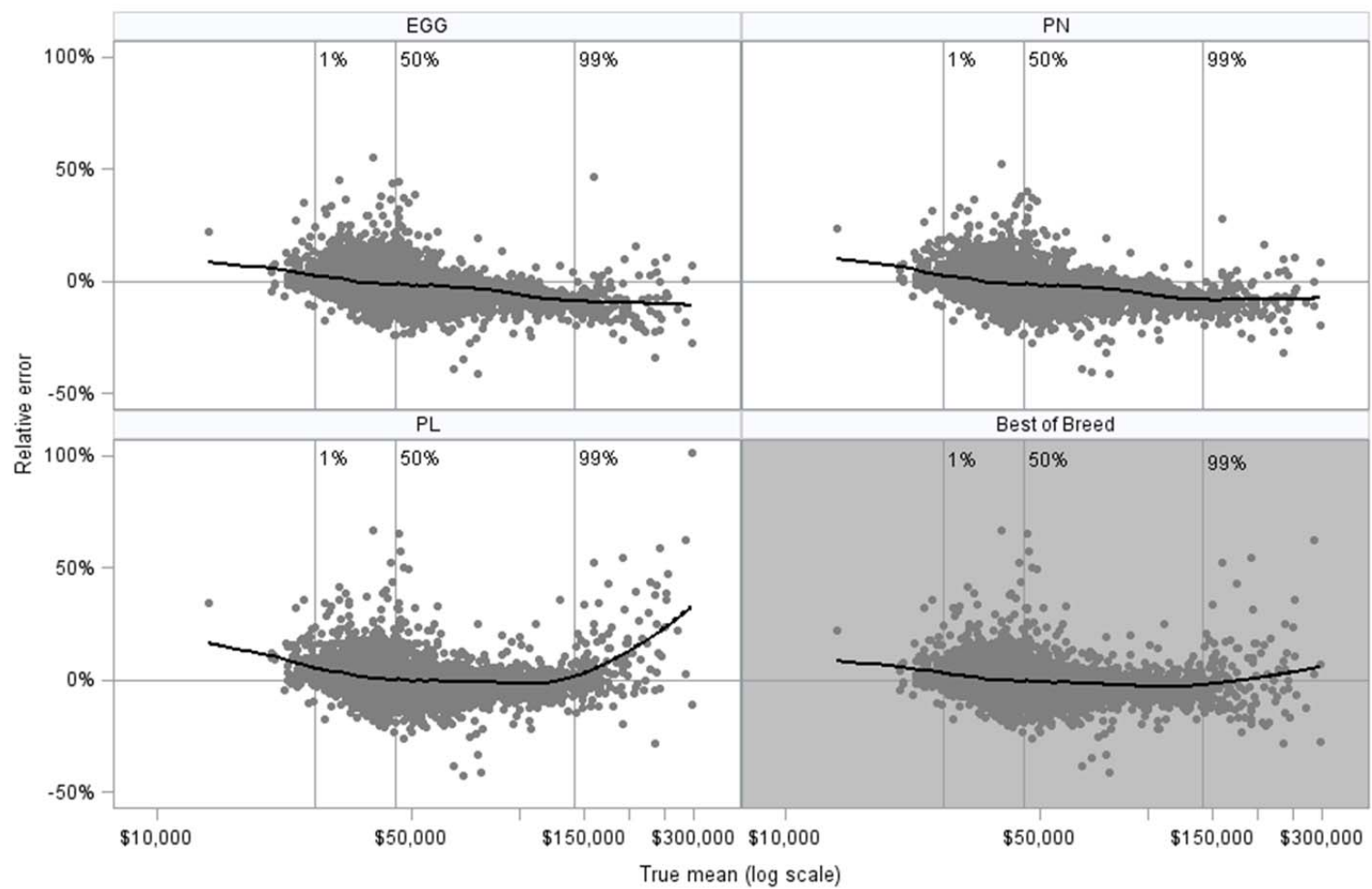

Figure 2. Relative error of estimated mean household income for U.S. school districts in 2000. The local regression curve estimates bias conditionally on the true mean.

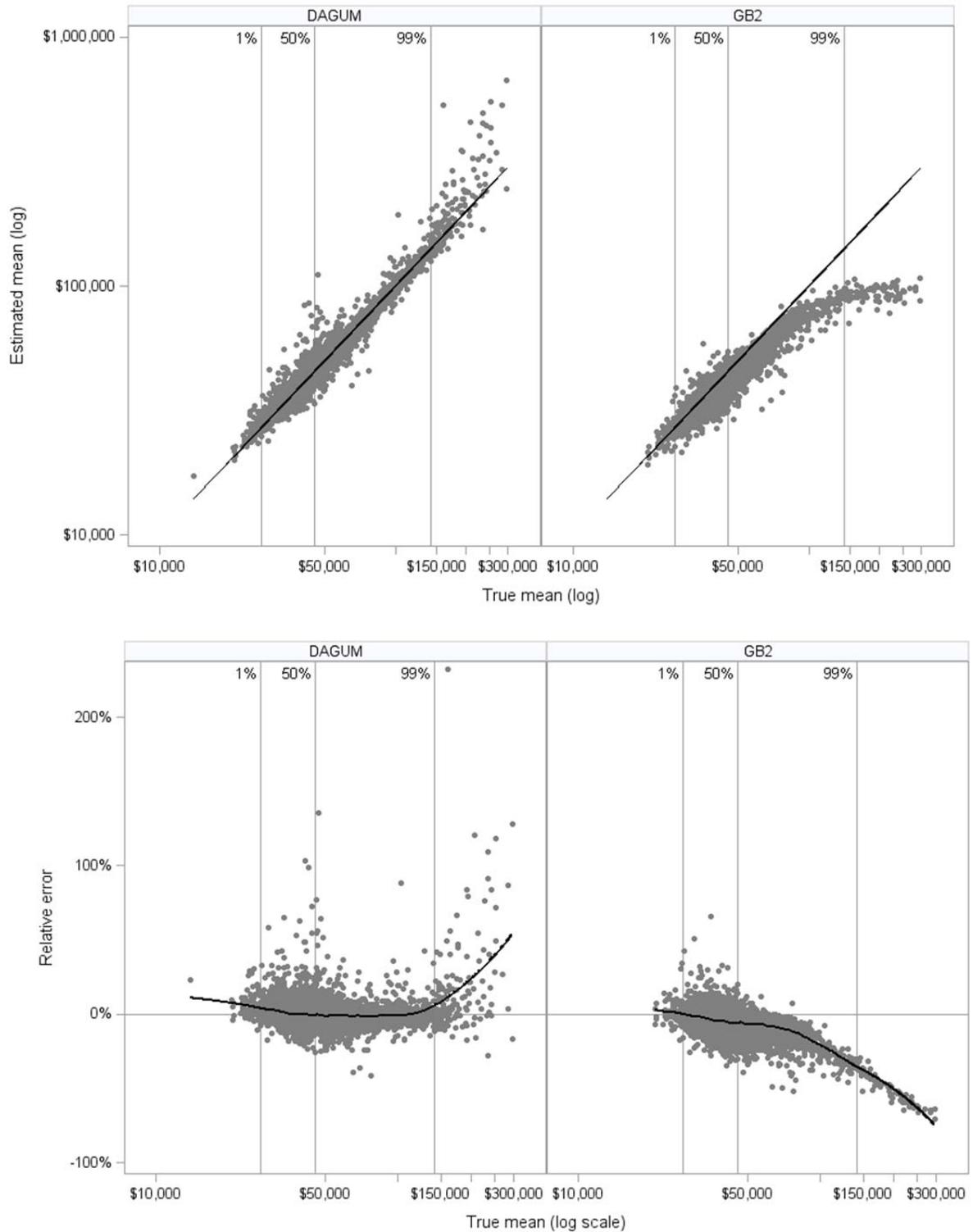

Figure 3. Top: Dagum and GB2 estimates of mean household income for U.S. school districts in 2000. Bottom: relative errors of those estimates. Not shown are the 16.5% of school districts for which the GB2 model does not estimate a finite mean.

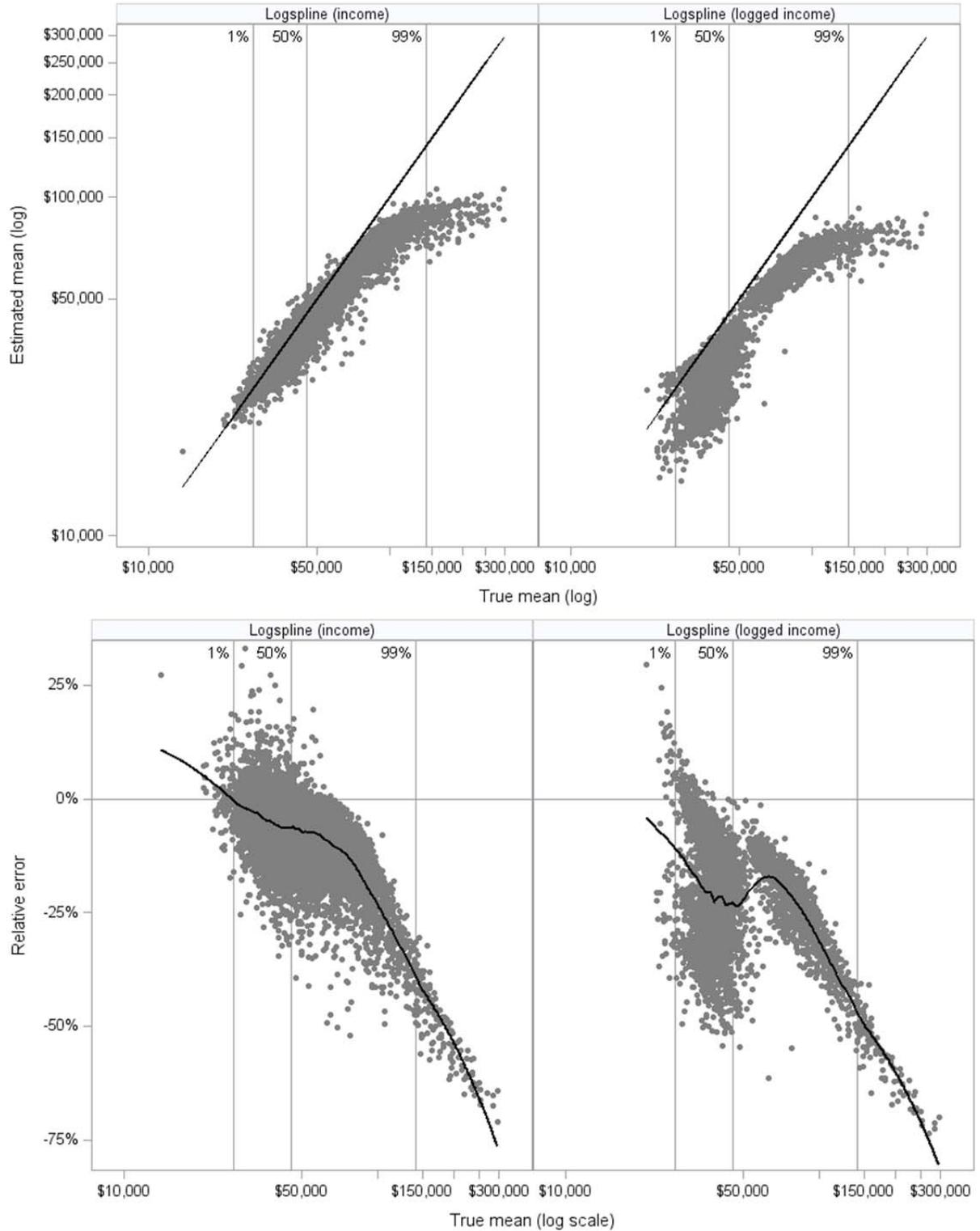

Figure 4. Top: Logspline estimates of mean household income for U.S. school districts in 2000, where the distribution is fit on the same scale as income (left), or on the scale of log income (right). Bottom: relative errors of those estimates. Not shown are the 51% of school districts for which the logged-income model does not estimate a finite mean.

# APPENDIX: THE %FIT_BINNED MACRO

## Version note

The macro was developed using SAS version 9.2 TS Level 2M3. It runs under that version, but it does not run under the later versions 9.3 TS Level 1M0 or Level 1M2.

## Definition

*%fit_binned* is a SAS macro with the following required keyword arguments:
- *data=* is the name of the dataset containing binned data
- *min=* and *max=* are the variables in *data* that represent the minimum and maximum value of each bin.
- *n=* is the name of the variable in *data* that represent the number of cases having values in each bin.
- *model=* is the name of the distribution to be fitted to the data. Possible values are *EGG*, *PN*, *PL*, and *best* (for best-of-breed).

The following arguments are optional:
- *print*=Y (the default) if the model estimates are to appear in the output window. Otherwise *print*=N.
- *estimates=* names the dataset to which the model estimates should be saved. By default the estimates are not saved to a permanent dataset.
- *id=* is the ID variable that distinguishes different school districts, countries, etc. This argument is only needed if *data* contains more than one school district, country, etc.

The output consists of the name of the fitted distribution (_DIST_), estimates of the parameters (mu, sigma, lambda), and estimates of the mean, variance, (sd), and coefficient of variation (cv).

## Example of use

The *%fit_binned* macro, and other macros on which it depends, are provided in the zipped archive *fit_binned.zip*. For purposes of this example, we assume that the files in *fit_binned.zip* have been copied to a folder called C:\SAS\fit_binned\.

Binned data for the two school districts in Table 1 are provided in the SAS datafile *two_districts.sas7bdat*. For purposes of this example we assume this datafile has been copied to a folder called C:\SAS\Data\.

Now the following code can be run as an example:

```
FILENAME binned 'C:\SAS\fit_binned\';
OPTIONS MAUTOSOURCE SASAUTOS=(SASAUTOS binned); /* Makes the
   %fit_binned macro available */
LIBNAME mydata 'C:\SAS\Data\'; /* Makes the district data available */

data mcnary;
 set mydata.two_districts;
 if district = "McNary";
run; /* Select the McNary district, then look at it */
proc print data=mcnary;
run;

/* Fit each of the models to the McNary district */
%fit_binned (data=mcnary, n=households, min=min, max=max, model=EGG);
%fit_binned (data=mcnary, n=households, min=min, max=max, model=PN);
%fit_binned (data=mcnary, n=households, min=min, max=max, model=PL);
%fit_binned (data=mcnary, n=households, min=min, max=max, model=best);

/* Fit the best-of-breed model to both districts, and copy the
   estimates to an output dataset */
%fit_binned (data=mydata.two_districts, estimates=best_for_each,
   n=households, min=min, max=max, model=best, id=district);
```